\input{aipcheck}

\def\alt{\stackrel{<}{\sim}}
\def\agt{\stackrel{>}{\sim}}
\def\ta{\tilde a}
\def\eslt{E_T^{\rm miss}}
\documentclass[
    ,final            
  ]
  {aipproc}

\layoutstyle{8x11single}

\begin{document}

\title{Physics Beyond the Standard Model\footnote{
To appear in the Proceedings of the Tenth Conference on the Intersections
of Particle and Nuclear Physics (CIPANP 2009), San Diego, California, 26-31 May, 2009.}}


\author{Howard Baer}{
  address={Dep't of Physics and Astronomy, University of Oklahoma, Norman, OK 73071}
}

\begin{abstract}
I present a brief overview of some exciting possibilities for physics
Beyond the Standard Model. I include short discussions of neutrino physics,
the strong CP problem and axions, GUTs, large and warped extra dimensions,
Little Higgs models and supersymmetry. 
The chances appear excellent that in the next few years--
as the LHC era gets underway-- 
data from a bevy of experiments will point the way to a new paradigm for the
laws of physics as we know them.
\end{abstract}

\maketitle


\section{The Standard Model}

The Standard Model (SM) emerged after many decades of hard work on the part of
experimentalists and theorists. It is predicated on just a few basic
principles\cite{altarelli}:
\begin{itemize}
\item postulate the local gauge symmetry of $SU(3)_C\times SU(2)_L\times U(1)_Y$,
\item input three generations of matter fermions: $Q_{Li}$, $u_{Ri}$, $d_{iR}$, $L_{Li}$ and $e_{Ri}$
with $i=1-3$, transforming under the appropriate gauge groups,
\item input a complex doublet of scalar fields along with an appropriate Higgs potential,
\item arrange all fields in the renormalizable lagrangian
${\cal L}={\cal L}_{gauge}+{\cal L}_{matter}+{\cal L}_{Higgs}+{\cal L}_{Yukawa}$ containing 19
free parameters.
\end{itemize}
While the Lagrangian is invariant under the gauge symmetry, the ground state of the theory, 
dictated by the Higgs potential, does not respect $SU(2)_L\times U(1)_Y$. The gauge symmetry is
spontaneously broken down the $SU(3)_C\times U(1)_{EM}$, giving mass to the weak gauge bosons, 
quarks and charged leptons. The neutrinos remain massless. The theory describes, often to 
exceedingly high precision, a vast array of data from high energy physics experiments. It is one of the
finest gems created by humanity, and explains a tremendous amount about the physical world we live in,
and has a major impact on astrophysics and cosmology as well. 

It is now clear from a variety of data that the limits of the descriptive power of the SM is being reached, and
we now need a new paradigm of physics Beyond the Standard Model (BSM).
From data, we have evidence for neutrino masses and mixings, the value of the neutron EDM, 
the matter-anti-matter asymmetry of the
universe, the existence of cold dark matter (CDM) in the universe and the existence of dark energy.
None of these can be (well) explained within the context of the SM.

On the theory side, quadratic divergences associated with the scalar sector require new physics
at or around the electroweak scale. Also, we have no guidance as to the origin of the generations, 
the quark and lepton masses and mixings or the origin of the gauge symmetries. We also need a 
consistent merging of the SM with a quantum mechanical theory of gravity.

On the positive side, data from existing experiments are already pointing the way to a new, more elegant 
paradigm for the laws of physics as we know them. In addition, data from experiments soon to operate, 
especially from the CERN LHC, should clinch the deal!

\section{The neutrino revolution}

While the SM includes only left-handed neutrino states, precluding massive neutrinos, the Davis-Bahcall
Homestake experiment was already yielding evidence for neutrino mass via its measurement of a
dearth of $\nu_e$s (relative to predicted rates from nuclear astrophysics calculations) 
coming from fusion reactions in the solar core. This ``solar neutrino problem'' was verified by the 
Gallex and Sage experiments. The likely solution was that neutrinos have mass, causing flavor oscillations.
In the case of the Sun, the MSW effect elegantly explained the neutrino deficit.
Atmospheric $\nu_\mu$ oscillations were observed by SuperK and confirmed by MINOS.
Meanwhile, the SNO experiment elegantly showed that the total solar neutrino emission rate was indeed in accord
with the Standard Solar Model, and KamLAND verified the mass gap and mixing angles needed\cite{nureview}.

We now have a picture of neutrino mass differences and mixing angles (see Fig. 1) which requires the addition of
gauge singlet right-hand neutrino states to the SM. The extremely light left-hand neutrinos, and
absence of right-hand neutrinos, can be elegantly explained by the see-saw mechanism, wherein an
ultra-heavy Majorana mass scale $M_N$ is introduced. The heavier the RHN mass scale, the lighter
the left-hand neutrino mass: hence the name see-saw. If a neutrino Dirac mass $m_D$ is induced via a
neutrino Yukawa coupling $f_\nu$, then $m_\nu\sim m_D^2/M_N$. The atmospheric mass gap is simply
explained for $M_N\sim 10^{15}$ GeV: very close to the Grand Unified Theory (GUT) mass scale.

\begin{figure}
  \includegraphics[height=.17\textheight]{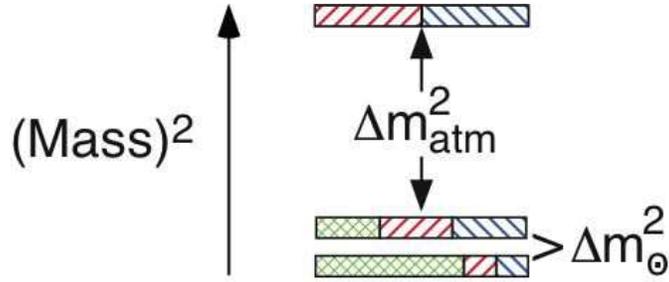}
  \caption{Neutrino mass splittings and mixings in the normal mass hierarchy}
\end{figure}

A bevy of neutrino experiments underway or in the planning stage should soon yield information on:
What are the absolute neutrino masses? What is the $\theta_{13}$ mixing angle? is the spectra
normal or inverted? Is there $CP$ violation in the $\nu$ sector? How does this all tie in
with GUTs and baryogenesis (possibly via leptogenesis)? 

\section{The strong $CP$ problem and axions} 

There is a saying that the laws of physics are a lot like sexual practices in Scandinavia:
{\it everything is allowed unless explicitly forbidden}! As an example, a term
${\cal L}\ni \frac{g_s^2\bar{\theta}}{32\pi^2}F_{\mu\nu}^A\tilde{F}^{\mu\nu A}$ is allowed
in the QCD Lagrangian, and in fact must be there according to 't Hooft's solution to the
$U(1)_A$ problem. However, such a term should give large contributions to the electric dipole moment of
the neutron. In contrast, such measurements restrict $\bar{\theta}\alt 10^{-11}$. Why $\bar{\theta}$
is so small is known as the strong $CP$ problem.

The Peccei-Quinn-Weinberg-Wilczek (PQWW) solution\cite{axreview} to the strong $CP$ problem is so elegant,
it would be a surprise to many if it were not true.
It hypothesizes a new global $U(1)_{PQ}$ symmetry which, when broken, leads to a pseudo-Goldstone boson:
the axion $a$. The PQWW solution allows the above Lagrangian term, but introduces the axion field through
PQ symmetry breaking at a scale $f_a\sim 10^9-10^{12}$ GeV. At temperatures $T\sim \Lambda_{QCD}$, an axion
potential turns on, and the axion field settles to a minimum where $a(x)\sim -\bar{\theta} f_a/N$, where 
$N$ is a model dependent factor. At the field minimum, the $CP$-violating Lagrangian term essentially vanishes,
and the strong $CP$ problem is solved. 

A consequence of the PQWW solution is that particle excitations of the axion field should exist. The axions
have extremely weak couplings-- suppressed by the PQ breaking scale. 
Their mass is expected to be in the $10^{-6}-10^{-3}$ eV range. While the decay $a\to\gamma\gamma$ can occur, 
the lifetime is far longer than the age of the universe. 
Axions are expected to be produced through field oscillations via the vacuum mis-alignment mechanism
in the early universe. Thus, axions are an excellent candidate for cold dark matter: see Fig. 2.

Axions can be searched for in experiments. The ADMX experiment looks for relic axions in a cryogenic microwave
cavity. The axion can interact with the $B$-field of the cavity and convert to a photon with energy 
equal to the axion mass. Only now are experiments such as ADMX reaching the sensitivity predicted by
theoretical models

\begin{figure}
  \includegraphics[height=.3\textheight]{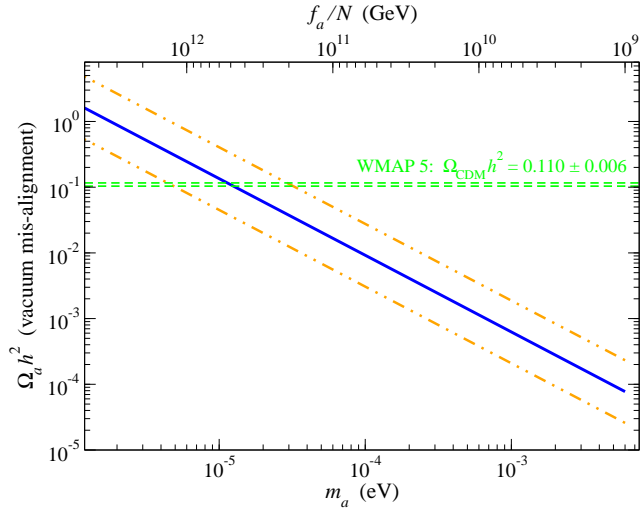}
  \caption{Relic density of axions versus $f_a/N$}
\end{figure}

\section{Grand unified theories (GUTs)}

Grand Unified Theories burst onto the particle physics scene in the mid-1970s with the $SU(5)$, 
$SO(10)$ and other models\cite{raby}. These theories unify the SM gauge groups, usually into a single gauge group, 
which can break to the SM group via spontaneous symmetry breaking. The original simple $SU(5)$ model
predicted proton decay at observable rates, and has since been excluded. Also, it soon became obvious that
simple GUT models suffered from the gauge herarchy problem, wherein the weak scale tends to blow up
to the GUT scale, unless tamed by extreme fine-tuning.  The combination of supersymmetry with GUT theories
in the early 1980s tamed the gauge hierarchy problem, and suppressed the $p$ lifetime. However, problems
remain, usually associated with the large Higgs representations needed for GUT symmetry breaking.

The big idea of (SUSY) GUTs is, IMO, almost certainly correct. 
This is exemplified by the celebrated unification of gauge couplings at $M_{GUT}$ in the SUSY SM.
However, all detailed SUSY GUT models
are almost certainly wrong. In recent years, GUTs had a revival, wherein the GUT theory could be
formulated in models with extra spacetime dimensions. Then, compactification of the extra dimensions on, say,
an orbifold can be used to break GUT symmetry instead of the Higgs mechanism. Examples of such models have been written 
down, and they can solve many problems endemic to 4-d SUSY GUTs. Ultimately, one wants to go to string theory,
and perhaps obtain many of the desired GUT properties from string compactification.

Nowadays, $SO(10)$ is probably more popular than $SU(5)$, in light of data on neutrino mass. In $SO(10)$,
all fermions of a single generation, plus the needed RHN field, can be combined neatly into the
16-dimensional spinor rep. Thus, $SO(10)$ unifies matter as well as gauge groups. In the simplest $SO(10)$ 
models,  Yukawa coupling of the third generation are also expected to unify at the GUT scale.

\section{Some exciting new models: large extra dimensions, warped extra dimensions and  Little Higgs}

\subsection{Large extra dimensions}

In the late 1990s, inspired by recent advances in superstring model 
building, Arkani-Hamed, Dvali and Dimopoulos proposed a simple model based on extra dimensions of
spacetime which contained some amazing properties\cite{add}. They proposed that nature was comprised of a
$d=4+n$ dimensional spacetime, referred to as the bulk. The Standard Model was proposed to
lie on a 3-brane embedded in the $d$-dimensional bulk. The fundamental mass scale in the bulk
was taken to be $M_*$, while the fundamental scale that we would observe in our 3-brane is the
usual Planck scale $M_P$.

If the $n$ extra dimensions were compactified {\it e.g.} on an $n$-dimensional torus of volume
$V_n$, then it turns out that our Planck scale $M_P$ is related to $M_*$ as $M_P^2=M_*^{n+2}V_n$.
A single extra dimension $n=1$ is ruled out by lack of deviations from inverse square gravity on the
solar system scale. But $n>1$ turns out to be allowed. In fact, if $V_n$ is large enough, then
$M_*$ can be as small as $m_{weak}$, offering a possible solution to the hierarchy problem
(of course, then one must understand why the Planck length $r_{Pl}$ is so much tinier than the 
compactification length $r_c$).

Since gravity is included in the model, upon compactification, one gets not only the SM, but also
KK excited gravitons. The massive KK gravitons can have mass of order the weak scale, but their
couplings to matter are still suppressed by a factor of $1/M_{Pl}$, so one might expect them to be 
irrelevant for collider searches. However, the mass splitting of the various KK gravitons is so tiny, 
that tremendous numbers of them may be produced at colliders. In fact, when summing the graviton emission
cross section over the entire spectrum of KK gravitons, the factors of $M_{Pl}$ cancel out, and
one finds possibly observable cross sections. Here is a model where quantum gravity effects are observable at
TeV scale colliders! And since the value of $M_*$ is so low, it might be possible in such a model to produce 
black holes at colliders. Such black holes would be expected to decay via Hawking radiation into a large
burst of particles. 

While the ADD model is ingenious and amusing, and gives rise to completely new effects
(visible massive graviton production and possibly BHs at colliders), I suspect few would
bet it as a likely option for BSM physics at the TeV scale.

\subsection{Warped extra dimensions} 

Another ingenious class of models based on extra dimensions is known as warped
extra dimensions, or Randall-Sundrum (RS) models\cite{rs}. The original RS model supposed the existence of
two $3-$branes embedded in a $5-d$ spacetime. Also assumed was a non-factorizable $5-d$ metric
of the form
\begin{equation}
ds^2=e^{-2k r_c\phi}\eta_{\mu\nu}dx^\mu dx^\nu +r_c^2d\phi^2
\end{equation}
where $\phi$ denotes distance along the extra dimension, assumed to compactify on an orbifold.
The solution to the $5-d$ Einstein equations shows that the spacetime is a slice of $AdS_5$,
and the exponential prefactor above is known as the {\it warp} factor. The mass scale on the
SM brane, $m_{weak}$, is related to the fundamental Planck scale by the warp factor
$$m_{weak}=e^{-kr_c\pi}M_{P} $$,
{\it i.e.} the weak scale emerges as the exponential suppression of the Planck scale via the warp factor!

In the RS model, the KK graviton zero mode remains massless, with couplings suppressed by factors of
$M_{P}$. However, the KK excited gravitons can have weak scale masses, with couplings suppressed
only by the weak scale! 
The spacing of the gravitons is of order the weak scale, and given by zeros of Bessel functions.
The RS massive gravitons can be produced as resonances at colliders. They decay into SM particles, and so
one anticipates an infinite sequence of resonances with spacings dictated by the Bessel functions!
Again, a model where quantum gravity effects are visible at accelerators!

\subsection{Little Higgs models}

Little Higgs (LH) models were proposed by Arkani-Hamed, Cohen and Georgi in 2001\cite{acg}. In these models, an enlarged
gauge symmetry is proposed. The Higgs field arises as a pseudo-Nambu-Goldstone boson from the
``collective symmetry breaking'' mechanism. The symmetry of the model guarantees that quadratic divergences
to the Higgs fields all cancel out at the one-loop level (two loop and higher quadratic divergences remain.)
The natural cut-off of the theory is taken to be $\Lambda\sim 10$ TeV, to avoid the ``little hierarchy
problem''.

All LH theories contain new particle states. To cancel off quadratic divergences due to top loops, a
new top partner fermion $T$ is needed. To cancel out gauge boson loops, new massive gauge bosons
$A_H$, $W^\pm_H$ and $W_H^0$ are expected. New scalars are also expected to cancel out Higgs self coupling loops.

A wide array of LH models were constructed. It was shown that this class of models had troubles 
matching precision electroweak data, unless a new parity, $T$-parity, was assumed\cite{cl}. While SM particles
were $t$-even, the new gauge bosons, some $t$-quark partners and scalars would be $t$-odd. Then
the new particles would enter precision electroweak observables only at loop level. And like
$R$-parity in SUSY, $t$-odd particles would have to be pair produced at colliders, they would have to
decay into other $t$-odd particles, and the lightest $t$-odd particle (LTP) would be absolutely stable.
Thus, the LTP could be a possible WIMP candidate for CDM.

A variety of collider searches for LH particles have been proposed. For instance, 
one might produce the $t$-quark partner $T$ at LHC via $pp\to T\bar{T}X$, followed by
$T\to tA_H$ decay. In this case, one would search for $t\bar{t}+\eslt$ events at LHC\cite{han}.

\section{Supersymmetry}

Supersymmetry-- like neutrino mass, the PQWW strong $CP$ solution, and GUTs-- is an idea almost too good
not to be true. The question is then, does SUSY have a connection to weak scale physics? Model building,
plus the measured values of the gauge couplings, the top-quark mass, and precision electroweak 
corrections all seem to point to: {\it yes}\cite{wss}. 
These data, matched against SUSY theory, seem to point to the Minimal Supersymmetric Standard Model (MSSM)
(or MSSM plus gauge singlets)
as being the correct effective theory of nature between the weak and GUT scales.
The truth will be revealed by experiment as the LHC era gets underway, since  weak scale
supersymmetry predicts a host of new matter states, many of which should be accessible to LHC searches.

At the LHC, the superpartners of the quarks and gluon-- the squarks and gluinos-- are expected to be produced
at observable rates, so long at sparticle masses are below about the 2-3 TeV scale\cite{lhc} (see Fig. 3). 
The MSSM provides
predictions for sparticle production cross sections and decay rates. These have been embedded into event
generators so that experimentalists can get an idea of what they are looking for. Different supersymmetric models
predict different sparticle masses and mixings, and hence different collider signatures. Thus, the sparticle
mass spectrum should bear the imprint of deeper underlying organizing principles. In the case of SUSY GUT
models, the sparticle mass spectrum would bear the imprint of GUT scale physics.

SUSY models are also compelling in that they may hold the solution to the dark matter problem.
SUSY theories contain a number of possible dark matter candidates. Most popular is the
lightest neutralino, which is a prototypical WIMP (weakly interacting massive particle)
candidate. SUSY models with neutralino dark matter usually predict too much dark matter in the
universe. Additional WIMP annihilations mechanisms are needed in the early universe to bring the
neutralino relic density into accord with WMAP measurements. In the minimal supergravity model
(mSUGRA or CMSSM), there are either 1. stau co-annihilation regions, 2. hyperbolic branch/focus point regions where
the neutralinos are mixed bino-higgsino particles, or 3. Higgs resonance annihilation regions. 
In region 1., neutralino dark matter direct detection (DD) and indirect detection (IDD) prospects are poor, 
while in region 2., DD and IDD prospects are excellent;
in region 3., IDD prespects from halo annihilations are good while DD and $\nu$-telescope prospects are poor\cite{bo}.

Gravitinos are also a possible SUSY CDM candidate, although these models suffer from severe constraints
from Big Bang Nucleosynthesis (BBN), since there would be late decays of sparticles into gravitinos plus other 
SM particles, causing breakup of the synthesized nuclei in the early universe.

A third alternative is axion dark matter. If the PQWW strong $CP$ solution is embedded into SUSY, then one
expects a supermultiplet containing an axion, a spin-${1\over 2}$ axino $\ta$ (and a spin zero saxion 
which is usually not so relevant for cosmology). If the axino is light-- its mass is model-dependent
and can range from the MeV to the multi-GeV range-- then dark matter can consist of three
components: cold axions from vacuum mis-alignment, warm axinos from neutralino (or other NLSP) decays, 
and thermally produced axinos (which qualify as CDM for $m_{\ta}\agt 100$ keV)\cite{steffen,axdm}.

A very compelling picture for BSM physics arises from $SO(10)$ SUSY GUT models. If one requires 
$t-b-\tau$ Yukawa coupling unification at $M_{GUT}$, then quantum corrections allow only a very constrained
sparticle mass spectrum: first/second generations scalars in the 10 TeV regime (beyond LHC reach),
third generation scalars and heavy Higgs scalars in the few TeV regime, and gluinos around 300-500 GeV 
(with a lightest neutralino around 50 GeV)\cite{bkss,raby2}. 
The light gluinos have a huge production cross section at LHC,
and should be easily discovered with less than 1 fb$^{-1}$ of integrated luminosity: they produce
a distinctive OS dimuon mass edge around 50-80 GeV, that serves as a smoking gun signature\cite{early}.
Since the neutralino is nearly pure bino, and scalars are extremely heavy, the neutralino relic density
comes in around $10^2-10^4$ times larger than WMAP measured value. 
However, if we solve the strong $CP$ problem via PQWW,
then the neutralinos can decay into MeV scale axinos, reducing the CDM relic density by factors of 
$10^3-10^5$! Detailed calculations actually favor the bulk of DM to be composed of axions, with a small
admixture of warm and cold axinos. This scenario is very compelling, and allows one to solve the gravitino 
problem (the gravitino mass is expected to be of order the heavy scalars $\sim 10$ TeV), and one obtains a
high enough re-heat temperature in the universe to at least explain baryogenesis via non-thermal 
leptogenesis\cite{bhkss}.


\begin{figure}
  \includegraphics[height=.3\textheight]{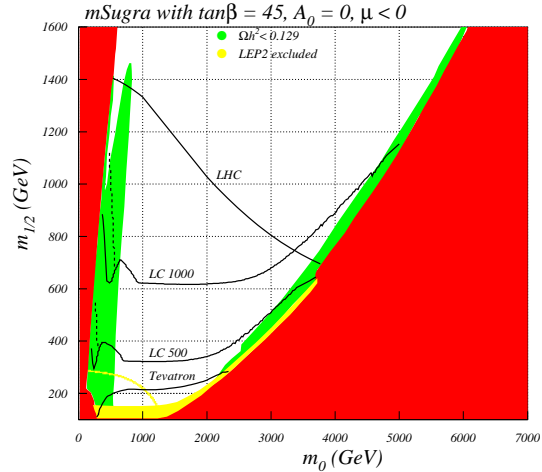}
  \caption{Sparticle reach of all colliders with relic density}
\end{figure}


\begin{theacknowledgments}
I thank Marvin Marshak for organizing an excellent conference!
\end{theacknowledgments}



\bibliographystyle{aipproc}   

\bibliography{sample}

\begin{thebibliography}{9}
%
\bibitem{altarelli} For a recent review of SM physics, see {\it e.g.} G. Altarelli, arXiv:0804.4147 (2008).
%
\bibitem{nureview} See {\it e.g.} B. Kayser, arXiv:0804.1497 (2008).
%
\bibitem{axreview} See {\it e.g.} J. E. Kim and G. Carosi, arXiv:0807.3125 (2008). 
%
\bibitem{raby} For a recent review, see S. Raby, Eur.\ Phys.\ J.\  C {\bf 59}, 223 (2009)
 
%
\bibitem{add} N. Arkani-Hamed, G. Dvali and S. Dimopoulos, Phys.\ Lett.\  B {\bf 429} (1998) 263. 
%
\bibitem{rs} L. Randall and R. Sundrum, Phys.\ Rev.\ Lett.\  {\bf 83} (1999) 3370.
%
\bibitem{acg} N. Arkani-Hamed, A. Cohen and H. Georgi, Phys.\ Lett.\  B {\bf 513} (2001) 232.
%
\bibitem{cl} H. C. Cheng and I. Low, JHEP {\bf 0309} (2003) 051.
%
\bibitem{han} T. Han, R. Mahbubani, D. Walker and L. T. Wang, JHEP {\bf 0905} (2009) 117. 
%
\bibitem{wss} For text book reviews, see, H.~Baer and X.~Tata, {\it Weak
Scale Supersymmetry: From Superfields to Scattering Events}, (Cambridge
University Press, 2006); M. Drees, R. Godbole and P. Roy,{\it Theory and
Phenomenology of Sparticles}, (World Scientific, 2004).
%
\bibitem{lhc} H. Baer, A. Belyaev, T. Krupovnickas and X. Tata, JHEP {\bf 0402} (2004) 007. 
%
\bibitem{bo} H. Baer and  J. O'Farrill, JCAP {\bf 0404} (2004) 005;
H. Baer, A. Belyaev, T. Krupovnickas and J. O'Farrill, 
JCAP {\bf 0408} (2004) 005; H. Baer, E. K. Park and X. Tata,
arXiv:0903.0555 (2009). 
%
\bibitem{steffen} F. Steffen, Eur.\ Phys.\ J.\  C {\bf 59} (2009) 557.
%
\bibitem{axdm} H. Baer, A. Box and H. Summy, arXiv:0906.2595 (2009).
%
\bibitem{bkss} H. Baer, S. Kraml, S. Sekmen and H. Summy, JHEP {\bf 0803} (2008) 056.
%
\bibitem{raby2} T.~Blazek, R.~Dermisek and S.~Raby, Phys.\ Rev.\  D {\bf 65} (2002) 115004.
%
\bibitem{early} H. Baer, H. Prosper and H. Summy, Phys.\ Rev.\  D {\bf 77} (2008) 055017;
H. Baer, A. Lessa and H. Summy, Phys. lett. {\bf B674} (2009) 49;
H. Baer, V. Barger, A. Lessa and X. Tata, arXiv:0907.1922 (2009).
%
\bibitem{bhkss} H. Baer and H. Summy, Phys. Lett. {\bf B666} (2008) 5;
H. Baer, M. Haider, S. Kraml,  S. Sekmen and H. Summy, JCAP {\bf 0902} (2009) 002.
%
\end{thebibliography}

\IfFileExists{\jobname.bbl}{}
 {\typeout{}
  \typeout{******************************************}
  \typeout{** Please run "bibtex \jobname" to optain}
  \typeout{** the bibliography and then re-run LaTeX}
  \typeout{** twice to fix the references!}
  \typeout{******************************************}
  \typeout{}
 }


\end{document}